\newcommand*{\wn}{cm$^{-1}$}
\def\apjl{Astroph.\ J.\ Lett.\ }

\def\plb{Phys.\  Lett.\ B}

\def\jms{J. Mol.\ Spectrosc.\ }

\def\aa{Astron.\ Astrophys.\  }
\def\aass{Astron.\ Astrophys.\ Suppl.\ Ser.\ }

\documentclass[aps,prl,twocolumn,superscriptaddress,showpacs]{revtex4}
\usepackage{graphics}
\usepackage{verbatim}
\usepackage{hyperref}
\usepackage{url}
\usepackage{amsmath}
\usepackage{natbib}

\begin{document}

\title{
Limits on a Gravitational Field Dependence of the Proton--Electron Mass Ratio
from H$_2$ in White Dwarf Stars}

\author{J. Bagdonaite}
\affiliation{Department of Physics and Astronomy, and LaserLaB, VU University, De Boelelaan 1081, 1081 HV Amsterdam, The Netherlands}
\author{E. J. Salumbides}
\affiliation{Department of Physics and Astronomy, and LaserLaB, VU University, De Boelelaan 1081, 1081 HV Amsterdam, The Netherlands}
\affiliation{Department of Physics, University of San Carlos, Cebu City 6000, Philippines}
\author{S. P. Preval}
\affiliation{Department of Physics and Astronomy, University of Leicester, University Road, Leicester LEI 7RH, United Kingdom}
\author{M. A. Barstow}
\affiliation{Department of Physics and Astronomy, University of Leicester, University Road, Leicester LEI 7RH, United Kingdom}
\author{J. D. Barrow}
\affiliation{DAMTP, Centre for Mathematical Sciences, University of Cambridge, Cambridge CB3 0WA, United Kingdom}
\author{M. T. Murphy}
\affiliation{Centre for Astrophysics and Supercomputing, Swinburne University of Technology, Melbourne, Victoria 3122, Australia}
\author{W. Ubachs}
\affiliation{Department of Physics and Astronomy, and LaserLaB, VU University, De Boelelaan 1081, 1081 HV Amsterdam, The Netherlands}

\date{\today}

\begin{abstract}

Spectra of molecular hydrogen (H$_2$) are employed to search for a possible proton-to-electron mass ratio ($\mu$) dependence on gravity. The Lyman transitions of H$_2$, observed with the Hubble Space Telescope towards white dwarf stars that underwent a gravitational collapse, are compared to accurate laboratory spectra taking into account the high temperature conditions ($T \sim 13\,000$~K) of their photospheres. We derive sensitivity coefficients $K_i$ which define how the individual H$_2$ transitions shift due to $\mu$-dependence. The spectrum of white dwarf star GD133 yields a $\Delta\mu/\mu$ constraint of $(-2.7\pm4.7_{\rm stat}\pm 0.2_{\rm sys})\times10^{-5}$ for a local environment of a gravitational potential $\phi\sim10^4\ \phi_\textrm{Earth}$, while that of G29$-$38 yields $\Delta\mu/\mu=(-5.8\pm3.8_{\rm stat}\pm 0.3_{\rm sys})\times10^{-5}$ for a potential of $2 \times 10^4$ $\phi_\textrm{Earth}$.

\end{abstract}


\maketitle

Theories of high-energy physics with a non-unique vacuum state, that
invoke extra dimensions, or contain new light scalar fields can permit or
require space-time variations of the fundamental low-energy ``constants'' of
nature \cite{Bekenstein1982, Uzan2011}. Small time-variations of
non-gravitational constants have negligible effects on the expansion
dynamics of the universe but have potentially observable effects on
astronomical spectra. Self-consistent scalar-tensor theories for the
variation of these constants (analogous to Brans-Dicke theory \cite{BD1961}
for a varying gravitation ``constant'', $G$) are needed to evaluate
their full cosmological consequences. Theoretical studies have focused on a
varying fine-structure constant $\alpha $, which is simplest to
develop because of its gauge symmetry \cite{Sandvik2002, Barrow2012}, and a varying proton-electron mass ratio $\mu =m_{p}/m_{e}$, \cite{Barrow2005a, Scoccola2008}.
Scaling arguments have been used to relate changes in $\alpha $, to changes in
$\mu $\ using the internal structure of the standard model, including
supersymmetry \cite{Calmet2002}. Typically (in the absence of
unusual cancellations involving the rates of change of $\alpha $, and the
supersymmetry-breaking and grand unification energy scales), they predict
that changes in $\mu $\ at low energies should be about an order of magnitude
greater than those in $\alpha $. However, high-redshift cosmological bounds on $\mu $\ variation are
expected to be weaker than those from laboratory tests of the equivalence
principle \cite{Barrow2005a}. Indications of possible variations
of $\alpha $\ in time~\cite{Webb1999} and space \cite{Webb2011} and
time variations in $\mu $~\cite{Reinhold2006} have been reported.
Systematic investigations of the spectra of cold H$_{2}$ towards
quasar sources have now produced a constraint on $\mu $-variation
over cosmological time scales yielding $\Delta \mu /\mu <1\times 10^{-5}$ at redshifts $z=2-3.5$, corresponding to look-back times of
10-12 Gyr \cite{Malec2010, Bagdonaite2014}.

Besides dependencies on cosmological scales, the couplings between
light scalar fields and other fields can generate dependencies of coupling
strengths on the local matter density~\cite{Sandvik2002,Khoury2004}, or on
local gravitational fields~\cite{Magueijo2002,Flambaum2008}. Such couplings violate the Einstein equivalence principle that is fundamental to General Relativity \cite{BD1961,Will2014}. The
gravitational potential at distance $R$ from an object of mass
$M$ is commonly expressed in dimensionless units of $
\phi =GM/(Rc^{2})$. A number of studies have been performed using
ultrastable lasers and atomic clocks exploiting the eccentricity of the
Earth's orbit~\cite{Ferrell2007,Fortier2007,Blatt2008,Shaw2008} causing
sinusoidal changes of $\Delta \phi =3\times 10^{-10}$. Recently, a
spectroscopic study of Fe\,V and Ni\,V ions in the local environment of the
photosphere of a white dwarf was employed to assess the dependence of $
\alpha $\ in a strong gravitational field ($\phi =4.9\times 10^{-5}$)~\cite{Berengut2013}. In the present study we use the spectrum of
molecular hydrogen in the photosphere of two white dwarfs, GD133 (WD\,1116$+$026) and G29$-$38 (WD\,2326$+$049), obtained with the Cosmic Origins
Spectrograph on the Hubble Space Telescope~\cite{Xu2013}, to probe a possible
dependence of $\mu $ on a gravitational potential that is $\sim
10^{4}$ times stronger than its value at the Earth's surface (which
is actually dominated by the contribution from the Sun's potential).

\begin{figure*}
\resizebox{0.95
\textwidth}{!}{\includegraphics{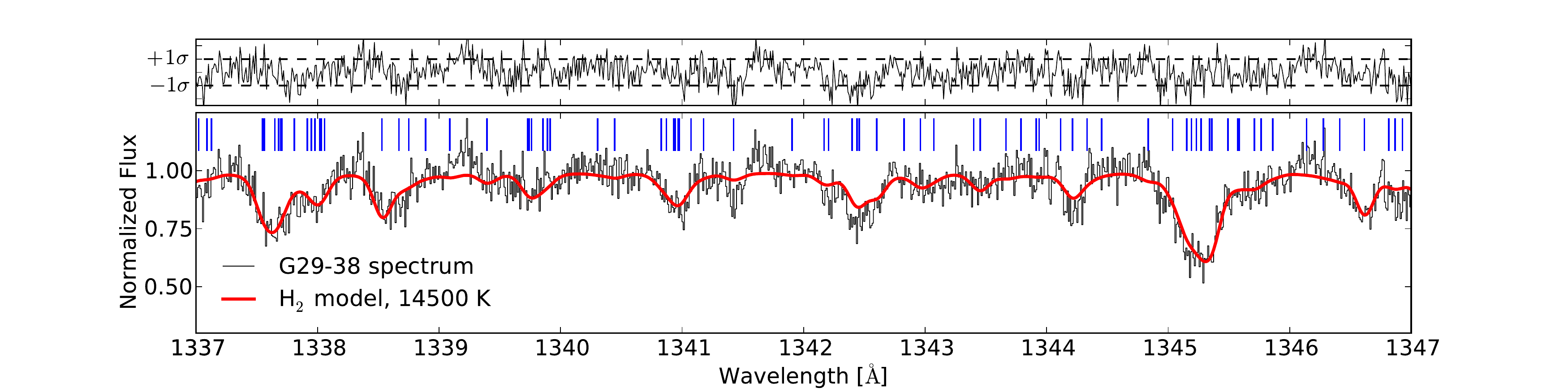}}
\caption{(Color online) Part of the white dwarf G29$-$38 spectrum in the range between 1337\,\AA\ and 1347\,\AA. Positions of the $B^1\Sigma_u^+$\,--\,$X^1\Sigma_g^+$ Lyman band H$_2$ transitions falling within this window are indicated with (blue) sticks. Overplotted (in red) on the spectrum is an absorption model based on the indicated transitions while the corresponding normalized residuals are shown at the top. To create the model, a set of fixed parameters known from molecular physics (transition rest-wavelength, $\lambda_i$, oscillator strength, $f_{ik}$, damping parameter, $\Gamma_i$) is combined with a partition function, $P_{v,J}(T)$, which defines level population at a temperature $T$. Even though this model includes many transitions, fitting it to the data requires only 4 free parameters: total column density $N$, redshift $z$, linewidth $b$, and temperature $T$. The total fitted spectrum extends from 1310.7\,\AA \ to 1411.7\,\AA \ \cite{Supp}.}
\label{fig1}
\end{figure*}

In Fig.~\ref{fig1} an overview of the H$_2$ absorption lines in the G29$-$38 photosphere is shown for the wavelength range 1337--1347 \AA. The total spectrum covers wavelengths from 1144 to 1444 \AA. The data of both G29$-$38 and GD133 were retrieved from the Hubble Space Telescope archive~\footnote{HST-archive, Cycle 18, program 12290, PI M. Jura.}. The individual exposures (3 of G29$-$38 and 5 of GD133) were rebinned to a common wavelength scale and combined using the same techniques as in \cite{Malec2010,Bagdonaite2014}. For both stars, lines pertaining to the $B^1\Sigma_u^+$\,--\,$X^1\Sigma_g^+$ Lyman band are solidly detected in the range 1298\,--\,1444\,\AA~at a signal-to-noise ratio of $\sim$15. The $C^1\Pi_u$\,--\,$X^1\Sigma_g^+$ Werner band transitions fall in the range 1144\,--\,1290\,\AA~at a lower signal-to-noise ratio of $\sim$5 and are only weakly detected and thus are not considered in the present analysis. Due to the high temperature in the photosphere, the observed H$_2$ lines are from multiple vibrationally -- and rotationally -- excited levels of the ground electronic state. The most intense H$_2$ Lyman transitions involve the $B-X$ ($v',v''$) bands for $v'=0-2$ and $v''=1-5$ vibrational levels with the highest population in the $J''=8$ level (at $T=13\,000$ K).

The laboratory wavelengths are derived from combination differences using level energies in the $B^1\Sigma^+_u, v', J'$ states from Refs.~\cite{Bailly2010, Abgrall1993}.
The  $X^1\Sigma^+_g$ ground state level energies used in the derivation are from \emph{ab initio} calculations including relativistic and quantum electrodynamical effects~\cite{Komasa2011}, with estimated uncertainties better than 0.001~\wn, which were tested in metrology laser experiments~\cite{Salumbides2011,Dickenson2013}.
The most accurate transition wavelengths are those derived from Ref.~\cite{Bailly2010} (for $J'<14$) with relative accuracies at the $10^{-8}$ level, while those derived from Ref.~\cite{Abgrall1993}, for higher $J'$ quantum numbers, exhibit relative accuracies of $10^{-6}$. When the level energies from Ref.~\cite{Abgrall1993} are used, an energy correction for each band is applied (typically 0.04-0.06 \wn) based on the comparison of Refs.~\cite{Bailly2010, Abgrall1993} at low $J$ quantum numbers.
The most intense transitions are listed in Table~\ref{Lyman}, and the complete list involves around 1500 lines~\cite{Salumbides}.

\begin{table}
\caption{The most intense H$_2$ transitions observed in GD133 and G29$-$38 spectra.
Wavelengths in \AA, with uncertainties in between parentheses given in units of the last digit. The calculated sensitivity coefficients $K_i$ and the oscillator strengths $f_{ik}$ are listed in the last two columns.}
\label{Lyman}
\begin{tabular}{c@{\hspace{10pt}}c@{\hspace{10pt}}c@{\hspace{10pt}}r@{.}l@{\hspace{10pt}}l@{\hspace{10pt}}c}
%B X trans wavelength(uncertainty) K_i
\toprule
$B, v'$ & $X, v''$ & Transition & \multicolumn{2}{c}{Wavelength} & $K_i$ & $f_{ik}$\\
\colrule
0	&3	&R(9)	&1\,313&376\,43\,(2)	&-0.106 &0.0494 \\
0	&3	&P(9)	&1\,324&595\,01\,(2)	&-0.115 &0.0538 \\
0	&3	&P(11)	&1\,345&177\,88\,(2)	&-0.129 &0.0508 \\
0	&4	&R(7)	&1\,356&487\,60\,(2)	&-0.114 &0.0821 \\
0	&4	&R(9)	&1\,371&422\,41\,(2)	&-0.125 &0.0816 \\
0	&4	&R(11)	&1\,389&593\,79\,(2)	&-0.138 &0.0816 \\
0	&4	&R(13)	&1\,410&648\,(1)	    &-0.152 &0.0821 \\
0	&4	&P(9)	&1\,383&659\,16\,(2)	&-0.134 &0.0739 \\
0	&4	&P(11)	&1\,403&982\,60\,(2)	&-0.148 &0.0765 \\
0	&4	&P(13)	&1\,427&013\,40\,(2)	&-0.163 &0.0793 \\
\botrule
\end{tabular}
\end{table}

\begin{figure}
\resizebox{0.44\textwidth}{!}{\includegraphics{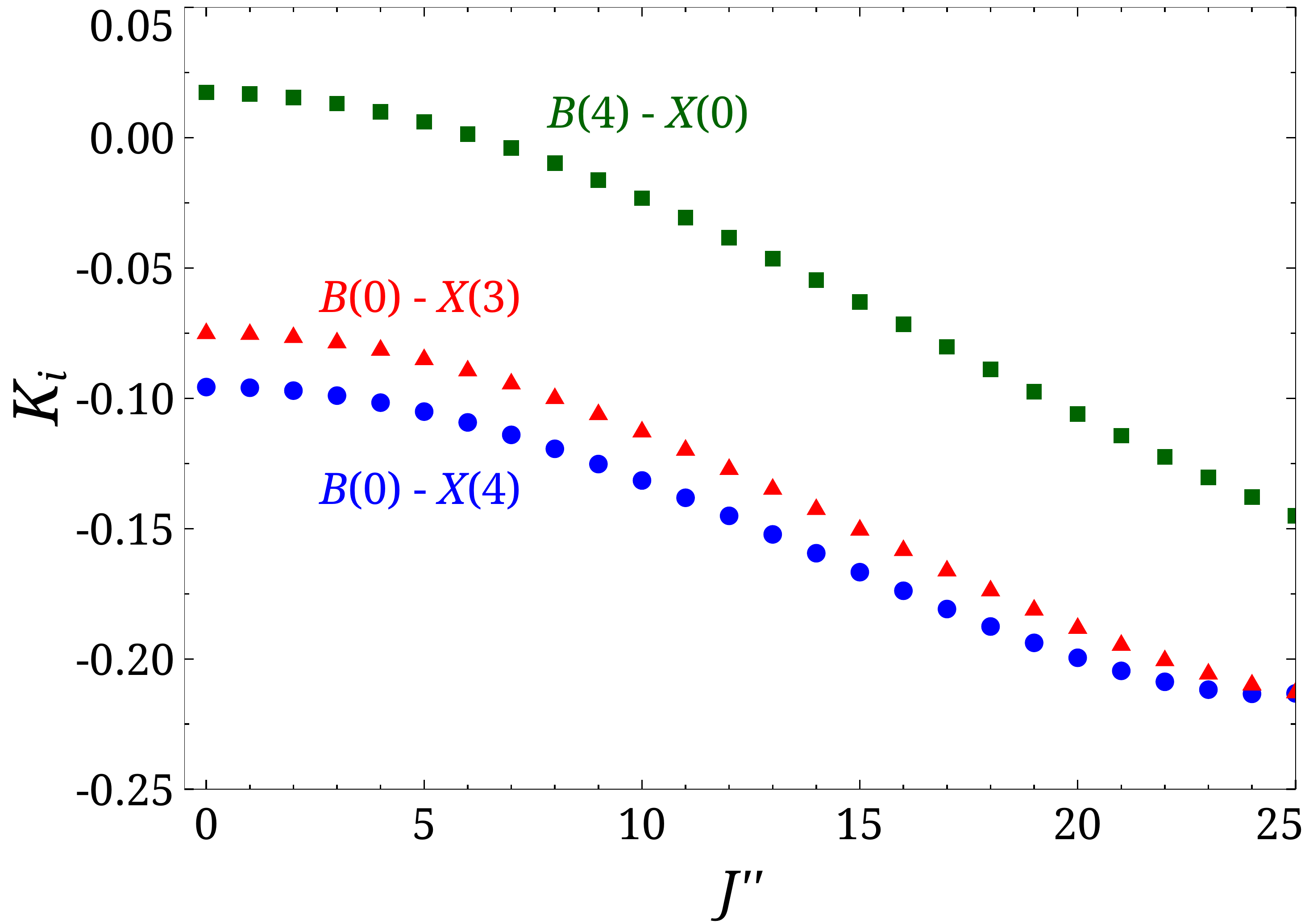}}
\caption{(Color online) $K_i$-coefficients for $R$-branch of the $B(0)-X(4)$ and $B(0)-X(3)$ Lyman bands which include the most intense H$_2$ transitions in GD133 and GD39$-$38 spectra. 
The $K_i$ values for the Lyman $B(4)-X(0)$ band, where low $J''<5$ transitions are used in probing $\mu$-variation in quasar absorption systems~\cite{Ubachs2007} are plotted for comparison.}
\label{K_i}
\end{figure}

Sensitivity coefficients $K_i$ due to a variation in $\mu$ were calculated for each transition $i$ using a semi-empirical method based on the experimentally-determined level energies.
The coefficient $K_i$ is separated into electronic ($K_\mathrm{el}$), vibrational ($K_\mathrm{vib}$) and rotational ($K_\mathrm{rot}$) contributions, which are calculated via
\begin{equation}
K_i=\frac{d\ln\lambda_i}{d\ln\mu}=-\frac{\mu}{E_B-E_X}(\frac{dE_B}{d\mu} - \frac{dE_X}{d\mu}),
\end{equation}
with $dE_{B,X}/d\mu$ related to $dE_{B,X}/dv$ and $dE_{B,X}/dJ$ to separate the $K_\mathrm{vib}$ and $K_\mathrm{rot}$ contributions. In the framework of the Born--Oppenheimer approximation the $K_\mathrm{el}$ are set to zero.
The method is related to the Dunham approach~\cite{Ubachs2007}, but turns out to be more robust due to the elimination of correlations in the fitting of the parameters of the Dunham matrix. This effect is more problematic for Dunham representations of levels with higher values of quantum numbers $v, J$, where the dominant contribution of the higher-order terms in the expansion are susceptible to numerical errors. In contrast, the present method does not have this disadvantage and the accuracy of the $K_i$ values are just limited by the experimental data.
The uncertainty of the $K_i$-coefficients is estimated to be as good as $10^{-4}$ for the observed Lyman transitions, especially because for $B, v'<8$ there are no perturbations with the $C^1\Pi_u$ electronic state in the probed wavelength range~\cite{Ubachs2007}. 
The results of the present method were verified to agree with the Dunham approach~\cite{Ubachs2007} for low $v, J$.
$K_i$-coefficients for the strongest Lyman bands observed in both white dwarfs, $B(0)-X(4)$ and $B(0)-X(3)$, are plotted in Fig.~\ref{K_i}. For comparison, the $B(4)-X(0)$ Lyman band observed in quasar absorption studies are also plotted, showing the higher sensitivity of the $B(0)-X(4)$ band despite the same $|\Delta v| = 4$.

To analyze the white dwarf spectra we do not follow the common procedure~\cite{Malec2010,Bagdonaite2014} of assigning and fitting individual transitions of H$_2$. Since they are relatively weak and self-blended we fit them simultaneously over most of the range between 1298 and 1444\,\AA. We only exclude regions where blends with atomic species occur: the geo-coronal O\,I transitions at 1298\,--\,1310~\AA, the photospheric and interstellar atomic transitions listed in \cite{Xu2014}, and part of the spectrum at $>$1411.7~\AA \ where some previously unidentified atomic transitions have been found. The non-linear least squares Voigt profile fitting program VPFIT10.0~\footnote{Developed by R. F. Carswell et al.; available at \url{http://www.ast.cam.ac.uk/\string~rfc/vpfit.html}} is used to model the absorption spectrum of H$_2$. A Voigt profile represents an absorption lineshape involving Doppler broadening, due to thermal motion of the absorbing gas, and Lorentzian broadening arising from the finite lifetimes of the excited states, represented by the damping parameter $\Gamma_i$~\footnote{For the present list of H$_2$ transitions $\Gamma_i$ parameters correspond to the total radiative transition probabilities from H. Abgrall, E. Roueff, and I. Drira, \aass {\bf 141}, 297 (2000).}, convolved with an instrumental line spread function~\footnote{We use the COS/HST ``lifetime position=2'' instrumental profiles provided at: \url{http://www.stsci.edu/hst/cos/performance/spectral_resolution/}}.

\begin{figure}
\resizebox{0.45\textwidth}{!}{\includegraphics{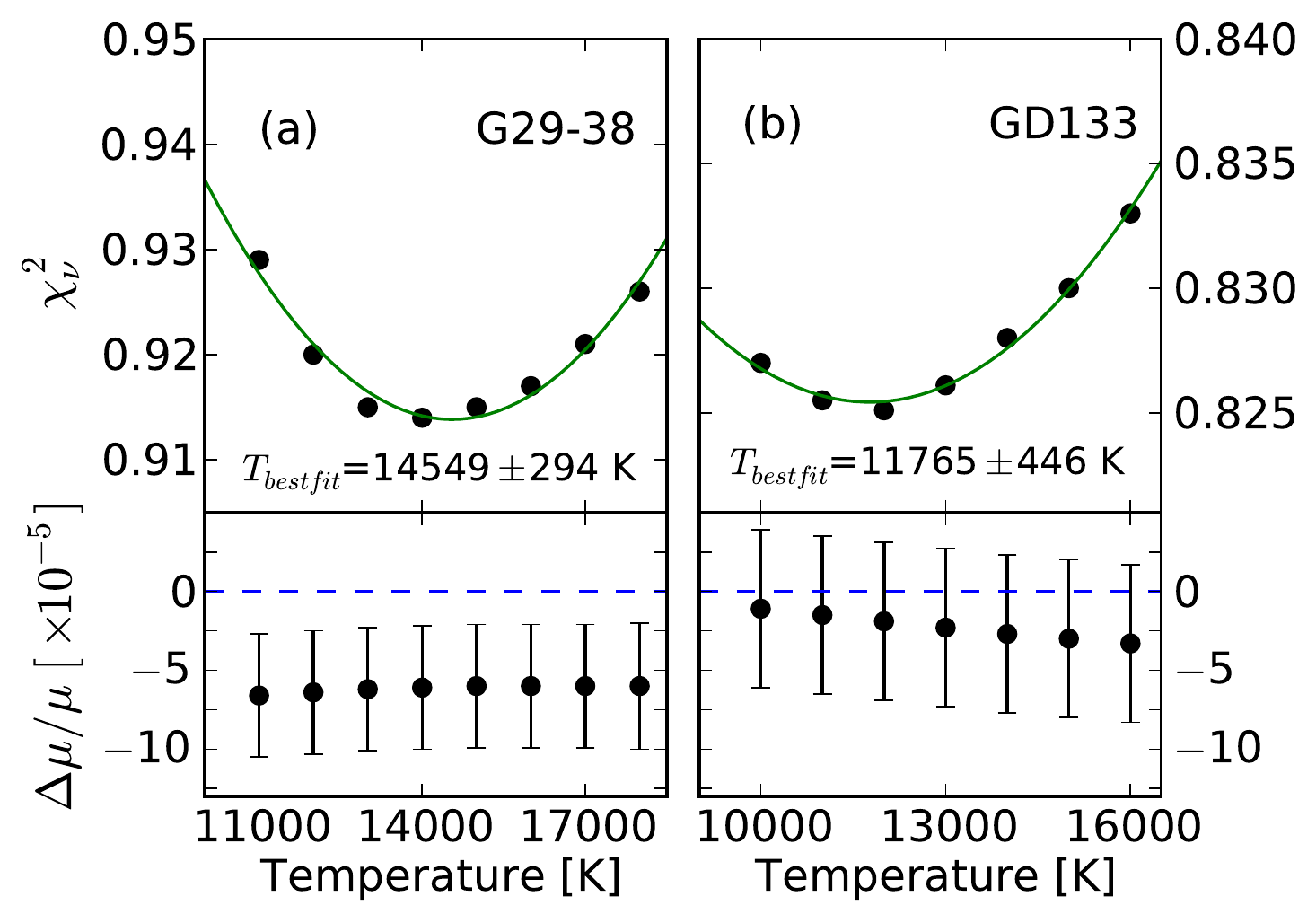}}
\caption{(Color online) (a) Top: Reduced $\chi^{2}$ from fitted H$_2$ models with varied temperatures, resulting in a value for temperature of H$_2$ in the G29$-$38 photosphere as indicated. Bottom: Measurements of $\Delta\mu/\mu$ invoking partition functions at different temperatures. (b) Same for GD133.}
\label{fig3}
\end{figure}

The intensities of the H$_2$ absorption lines are described by the product of the oscillator strength $f_{ik}$ and the normalized population of the ground ro-vibrational level, calculated from a partition function at a temperature $T$:
\begin{equation}
P_{v,J}(T) = \frac{g_n (2J+1) e^{\frac{-E_{v,J}}{kT}}}{\sum\limits_{v=0}^{v_\textrm{max}} \sum\limits_{J=0}^{J_\textrm{max}(v)} g_n (2J+1) e^{\frac{-E_{v,J}}{kT}}}
\label{partition-function}
\end{equation}
where $g_n$ is the nuclear statistical weight. Consequently, all lines probing odd-$J$ levels (ortho-H$_2$) benefit in relative strength from the 3:1 spin statistics ratio between ortho- and para-levels. Invoking this definition of line strengths leads to a model which essentially requires only 4 free parameters: total column density of the gas, $N$, redshift of the absorbing cloud, $z$, linewidth, $b$, and temperature $T$. Once the fit is optimized, we introduce an additional free parameter $\Delta\mu/\mu$ which allows for small relative line-shifts that are governed by the calculated sensitivity coefficients $K_i$:
\begin{equation}
 \frac {\lambda_i^{\mathrm{WD}}}{\lambda_i^0} =  (1+z_\mathrm{WD})(1+ \frac{\Delta \mu}{ \mu}K_i)
\label{redshift-1}
\end{equation}
where $\lambda_i^{\mathrm{WD}}$ represents the transition wavelength observed in the white dwarf spectra and $\lambda_i^0$ is a corresponding wavelength measured in the laboratory. Models of different temperatures were fitted to the data (Fig.~\ref{fig3}), resulting in a best-fit temperature of $T=(11\,800\pm450)$~K for GD133 and $T=(14\,500\pm300)$~K for G29$-$38. Displayed in Fig.~\ref{fig1} are fitting results of the G29$-$38 spectrum, where the model is based on a total of $\sim$870 H$_2$ transitions with the relative strengths defined for $T=14\,500$ K. The derived H$_2$ temperature is in good agreement with independent temperature determinations from Balmer-H lines for GD133 but differs for GD29$-$38~\cite{Koester2009}. For either star, measurements of $\Delta\mu/\mu$ are only slightly affected by the choice of temperature, as shown in Fig.~\ref{fig3}. The best-fit model of GD29$-$38 yields $\Delta\mu/\mu=(-6.1\pm3.9)\times 10^{-5}$, and the one of GD133 results in $\Delta\mu/\mu=(-1.8\pm5.0)\times 10^{-5}$. The adequacy of the fit is reflected by the reduced $\chi_{\nu}^2$ of $\sim0.8-0.9$, where the number of degrees of freedom $\nu$ is equal to 9633 for both spectra. A column density $\log [N/{\rm cm}^{-2}]  = (15.849\pm0.007)$, a linewidth $b = (14.55\pm0.58)$~km\,s$^{-1}$, and a redshift $z=0.0001820(10)$ were measured for GD133. For G29$-$38 the results are: $\log [N/{\rm cm}^{-2}] = (15.491\pm 0.005)$, $b = (18.65\pm 0.42)$~km\,s$^{-1}$, and  $z=0.0001360(8)$. The quoted widths $b$ are deconvolved from the instrument profile, $FWHM~\simeq~17$~km\,s$^{-1}$.

The measured redshifts of $z_{\rm{WD}}\sim10^{-4}$ are primarily determined by the gravitational redshift associated with the local potential in the white dwarf photospheres, with contribution from the proper motion of the objects and from the uncertainty of the absolute wavelength calibration of the COS instrument, amounting to $1/3$ of the measured redshift value. For a $\Delta\mu/\mu$-analysis, the relative wavelength calibration accuracy is of the utmost importance. If not taken into account, velocity distortions -- velocity shifts which change with wavelength -- may have a significant effect on $\Delta\mu/\mu$ measurements \cite{Bagdonaite2014}. We searched for such distortions by applying the `direct comparison method' \cite{Evans2013} to individual exposures against the combined spectra. The `direct comparison method' is a model-independent technique of comparing pairs of spectra in order to detect and correct for velocity shifts. No evidence for relative distortions between exposures was found, with 1-$\sigma$ limits of $\Delta v^\prime<$25\,m\,s$^{-1}$\,nm$^{-1}$ per exposure. Applying artificial distortions of $\Delta v^\prime/\sqrt{3}$ and $\Delta v^\prime/\sqrt{5}$ to the combined 3 exposures of G29$-$38 and the 5 of GD133 produces systematic shifts in $\Delta\mu/\mu$ of $\sigma_{\rm sys}=\pm 0.3$ and $\pm 0.2\times10^{-5}$, respectively. The same analysis also allows us to correct the combined spectra for small relative shifts ($<$0.2\,km\,s$^{-1}$) between individual exposures. The corrected spectra, plus the above estimate of systematic errors, provide our fiducial measurements: $\Delta\mu/\mu=(-5.8\pm3.8_{\rm stat}\pm 0.3_{\rm sys})\times10^{-5}$ for G29$-$38 and $(-2.7\pm4.7_{\rm stat}\pm 0.2_{\rm sys})\times10^{-5}$ for GD133.

The line broadenings of $15$ and $19$~km\,s$^{-1}$ are primarily determined by gas kinetics at the prevailing temperatures of $T=12\,000-14\,000$ K yielding $b_\textrm{therm}\sim 10$~km\,s$^{-1}$, with $b_\textrm{therm} = \sqrt{2kT/m}$, where $k$ is the Boltzmann constant, and $m$ the molecular mass. It can be estimated that an H$_2$ absorption cloud of 10 km depth in the photosphere of  GD133 would be subject to a ``gravitational width'' of $(\Delta\lambda/\lambda)_\textrm{grav} = 3.2 \times 10^{-8}$ or only $b_\textrm{grav} = 0.01$~km\,s$^{-1}$ with a similarly small estimate for G29$-$38.
From a photospheric model the maximum H$_2$ molecular density ($n$(H$_2$)/$n$(H)=10$^{-5}$) was found to coincide with a total material density of $\rho = 1.5\times10^{-7}$\,g\,cm$^{-3}$~\cite{Xu2014} amounting to $\sim$3 mbar for an H-atmosphere. This would translate into a broadening of $b_\textrm{col} < 0.001$~km\,s$^{-1}$.

Stark and Zeeman broadening effects on the H$_2$ lines are assumed to be small because the excited state $B^1\Sigma_u^+$ for the Lyman bands is of valence character and only weakly susceptible to external fields; no laboratory measurements of Stark and Zeeman effects have been reported for the molecular Lyman bands. G29$-$38 is a well-studied irregular pulsator, with time-evolving dominant periods of a few hundred seconds resulting in velocity shifts of atomic hydrogen transitions as large as 16.5~km\,s$^{-1}$~\cite{Thompson2003}. In the case of GD133 pulsations occur at a dominant period of 120 s and are much weaker than in G29$-$38~\cite{Silvotti2006}. In either case, the pulsation period is smaller than the exposure times that exceed 2000~s and, thus, the H$_2$ spectra in individual exposures and the combined spectra will be smeared out by this effect. 

The broadening effects outlined here mainly affect the resolution of the spectra and the accuracy of the constraint on $\Delta\mu/\mu$. They should not affect the symmetry of the H$_2$ absorption lines, and even if they did, the effect would be the same for all transitions and thus it is unlikely to mimic $\mu$-variation.

The above constraints on $\Delta\mu/\mu$ from the white dwarf spectra can be interpreted in terms of a dependence on a dimensionless gravitational potential~\cite{Magueijo2002,Flambaum2008}, $\Delta\phi$, where a Taylor series expansion up to the second order can be used to specifically probe strong field gravitational phenomena:
\begin{equation}
 \frac {\Delta\mu}{\mu} =  k^{(1)}_{\mu} \Delta\phi + k^{(2)}_{\mu} (\Delta\phi)^2
\label{Dmu-int}
\end{equation}
The linear term can be constrained most directly from a laboratory spectroscopic investigation of SF$_6$ molecules aimed to detect a temporal variation of the proton-to-electron mass ratio~\cite{Shelkovnikov2008}. We estimate from the results presented in Ref.~\cite{Shelkovnikov2008} that the seasonal difference amounts to $|\Delta\mu/\mu| < 10^{-13}$. Invoking a gravitational potential at the Earth's surface of $\phi_\textrm{Earth} = 0.98 \times 10^{-8}$ (due to the field produced by the Sun) and an Earth orbit eccentricity of $\epsilon = 0.0167$ leading to a 2.6\,\% effect on the difference in the potential between aphelion and perihelion in the current epoch, the laser spectroscopic experiment yields $k^{(1)}_{\mu} < 4 \times 10^{-4}$
\footnote{It is noted that most of the spectroscopic measurements of~\cite{Shelkovnikov2008} were fortuitously taken at the aphelion period (July 2005 and 2006), while some were recorded at perihelion (in 2004 and 2006), thus probing the maximum potential difference produced during the Earth's orbit.}.
Other Earth-based spectroscopic investigations (and combinations thereof) yield even tighter constraints of $k^{(1)}_{\mu} < (4.9 \pm 3.9) \times 10^{-5}$~\cite{Ferrell2007,Fortier2007}, and $k^{(1)}_{\mu} < (-1.3 \pm 1.7) \times 10^{-5}$~\cite{Blatt2008}, although with model-dependence. These results constrain the linear term, $k_\mu^{(1)}$ more than the present white dwarf study.

The analysis of H$_2$ spectral lines in the white dwarfs yield $|\Delta\mu/\mu| \lesssim 5 \times 10^{-5}$. The physical properties of GD133 correspond to a gravitational potential $\phi = 1.2 \times 10^{-4}$ in the photosphere at the white-dwarf surface, while that of G29$-$38 is $\phi = 1.9 \times 10^{-4}$. This delivers a constraint of $k^{(2)}_{\mu} < 1 \times 10^{3}$, which is several orders of magnitude more stringent than from the Earth-based experiments. This demonstrates that the high gravitational field conditions of white dwarfs (10,000 times that on the Earth's surface) is a sensitive probe to constrain $k^{(2)}_{\mu}$. Using the methods presented here, future studies of H$_2$ in the photospheres of white dwarfs should provide further information on the possible variation of the proton-to-electron mass ratio under conditions of strong gravitational fields.

This work was supported by the FOM-Program ``Broken Mirrors \& Drifting Constant'', Science and Technology Facilities Council, Templeton Foundation and Australian Research Council (DP110100866). Tyler Evans (Swinburne University of Technology) is thanked for calculating velocity shifts between individual exposures.


\begin{thebibliography}{}

\bibitem{Bekenstein1982} J. D. Bekenstein, \prd~{\bf 25}, 1527 (1982).
\bibitem{Uzan2011} J.-P. Uzan, Living Rev. Relativity~{\bf 14}, 2 (2011).
\bibitem{BD1961} C. Brans and R.H. Dicke, Phys. Rev.~{\bf 124}, 925 (1961).
\bibitem{Sandvik2002} H. B. Sandvik, J. D. Barrow, and J. Magueijo, \prl~{\bf 88}, 031302 (2002); J. D. Barrow, H. B. Sandvik, and J. Magueijo, \prd~{\bf 65}, 063504 (2002).
\bibitem{Barrow2012} J. D. Barrow and S. Z. W. Lip, \prd~{\bf 85}, 023514 (2012); J. D. Barrow and J. Magueijo, gr-qc/1406.1053.
\bibitem{Barrow2005a} J. D. Barrow, \prd~{\bf 71}, 083520 (2005); J. D. Barrow and J. Magueijo, \prd~{\bf 72}, 043521 (2005).
\bibitem{Scoccola2008} C. G. Scoccola, M. E. Mosquera, S. J. Landau and H. Vucetich, Ap. J.~{\bf 681}, 737 (2008).
\bibitem{Calmet2002} X. Calmet and H. Fritsch, Eur. J. Phys. C~{\bf 24}, 639 (2002); X. Calmet and H. Fritzsch, Europhys. Lett.~{\bf 76}, 1064 (2006).
\bibitem{Webb1999} J. K. Webb, V. V. Flambaum, C. W. Churchill, M. J. Drinkwater,
  and  J. D. Barrow, \prl~{\bf 82}, 884 (1999).
\bibitem{Webb2011} J. K. Webb, J. A. King, M. T. Murphy, V. V. Flambaum, R. F. Carswell, M. B. Bainbridge, \prl~{\bf 107}, 191101 (2011).
\bibitem{Reinhold2006} E. Reinhold, R. Buning, U. Hollenstein, A. Ivanchik, P. Petitjean, and W. Ubachs, \prl~{\bf 96}, 151101 (2006).
\bibitem{Malec2010} A. L. Malec, R. Buning, M. T. Murphy, N. Milutinovic, S. L. Ellison,
J. X. Prochaska, L. Kaper, J. Tumlinson, R. F. Carswell, and W. Ubachs,
 MNRAS~{\bf 403}, 1541 (2010); F. van Weerdenburg, M. T. Murphy, A. L. Malec, L. Kaper, and W. Ubachs, \prl~{\bf 106}, 180802 (2011).
\bibitem{Bagdonaite2014} J. Bagdonaite, W. Ubachs, M. T. Murphy, and J. B. Whitmore,
 \apj {\bf 782}, 10 (2014).
\bibitem{Khoury2004} J. Khoury and A. Weltman, \prl~{\bf 93}, 171104 (2004).
\bibitem{Magueijo2002} J. Magueijo, J. D. Barrow and H. Sandvik, \plb~{\bf 549}, 284 (2002).
\bibitem{Flambaum2008} V. V. Flambaum, and E. V. Shuryak, Nuclei and Mesoscopic Physics - WNMP 2007, AIPC Proceedings {\bf 995}, 1 (2008).
\bibitem{Will2014} C.~M. Will, Living Rev. Relativity {\bf 17}, 4 (2014).
\bibitem{Ferrell2007} S. J. Ferrell, A. Cing\"{o}z, A. Lapierre, A.-T. Nguyen, N. Leefer, D. Budker, V. V. Flambaum, S. K. Lamoreaux, and J. R. Torgerson, \pra~{\bf 76}, 062104 (2007).
\bibitem{Fortier2007} T. M. Fortier, N. Ashby, J. C. Bergquist, M. J. Delaney, S. A. Diddams, T. P. Heavner, L. Hollberg, W. M. Itano, S. R. Jefferts, K. Kim, F. Levi, L. Lorini, W. H. Oksay, T. E. Parker, J. Shirley, and J. E. Stalnaker, \prl~{\bf 98}, 070801 (2007).
\bibitem{Blatt2008} S. Blatt, A. D. Ludlow, G. K. Campbell, J. W. Thomsen, T. Zelevinsky, M. M. Boyd, J. Ye, X. Baillard, M. Fouch\'{e}, R. Le Targat, A. Brusch, P. Lemonde, M. Takamoto, F.-L. Hong, H. Katori, and V. V. Flambaum, \prl~{\bf 100}, 140801 (2008).
\bibitem{Shaw2008} J. D. Barrow and D. J. Shaw, Phys. Rev. D~{\bf 78},
067304 (2008).
\bibitem{Berengut2013} J. C. Berengut, V. V. Flambaum, A. Ong, J. K. Webb, J. D. Barrow,
  M. A. Barstow, S. P. Preval, and J. B. Holberg, \prl~{\bf 111}, 010801 (2013).
\bibitem{Xu2013} S. Xu, M. Jura, D. Koester, B. Klein, and B. Zuckerman,
 \apjl~{\bf 766}, L18 (2013).
\bibitem{Supp} Supplementary Material.
\bibitem{Bailly2010} D. Bailly, E. J. Salumbides, M. Vervloet, and W. Ubachs,
    Mol. Phys.~{\bf 108}, 827 (2010).
\bibitem{Abgrall1993} H. Abgrall, E. Roeff, F. Launay, J.-Y. Roncin, and J.-L. Subtil, \jms~{\bf 157}, 512 (1993).
\bibitem{Komasa2011} J. Komasa, K. Piszczatowski, G. \L ach, M. Przybytek, B. Jeziorski, and K. Pachucki, J. Chem. Theory Comput. {\bf 7}, 3105 (2011).
\bibitem{Salumbides2011} E. J. Salumbides, G. D. Dickenson, T. I. Ivanov, and W. Ubachs,
 \prl~{\bf 107}, 043005 (2011).
\bibitem{Dickenson2013} G. D. Dickenson, M. L. Niu, E. J. Salumbides, J. Komasa,
 K. S. E. Eikema, K. Pachucki, and W. Ubachs,
 \prl~{\bf 110}, 193601 (2013).
\bibitem{Salumbides} A full list of transition wavelengths, oscillator strengths and sensitivity coefficients $K_i$ will be published later by E. J. Salumbides et al.
\bibitem{Ubachs2007} W. Ubachs, R. Buning, K. S. E. Eikema, E. Reinhold,
 \jms~{\bf 241}, 155 (2007).
%\bibitem{Cline1970} D. Cline, and P.~M.~S. Lesser, Nucl. Instrum. and Meth. {\bf 82}, 291 (1970).
\bibitem{Xu2014} S. Xu, M. Jura, D. Koester, B. Klein, and B. Zuckerman,
 \apj~{\bf 783}, 79 (2014).
\bibitem{Koester2009} D. Koester, B. Voss, R. Napiwotzki, N. Christlieb, D. Homeier, T. Lisker, D. Reimers, and U. Heber,
 \aa~{\bf 505}, 441 (2009).
\bibitem{Evans2013} T.~M. Evans and M.~T. Murphy, \apj~{\bf 778}, 173 (2013).
\bibitem{Thompson2003} S.~E. Thompson, J.~C. Clemens, M.~H. van Kerkwijk, and D. Koester, \apj~{\bf 589}, 921 (2003).
\bibitem{Silvotti2006} R. Silvotti, M. Pavlov, G. Fontaine, T. Marsh, and V. Dhillon, Mem. S. A. It. {\bf 77}, 486 (2006).
\bibitem{Shelkovnikov2008} A. Shelkovnikov, R. J. Butcher, C. Chardonnet, and A. Amy-Klein, \prl~{\bf 100}, 150801 (2008).
\end{thebibliography}
\end{document}